\documentclass[preprint2]{aastex}
\usepackage{graphicx,times}             %for PS/EPS graphics inclusion, new
\usepackage{epstopdf}

\shorttitle{Abundance Boundaries of the Heavier neutron-capture Elements}
\shortauthors{Yang et al.}

\begin{document}


\title{Formation of the Abundance Boundaries of the Heavier Neutron-capture
Elements in Metal-poor Stars}


\author{Guochao Yang\altaffilmark{1}, Hongjie Li\altaffilmark{1,2}, Nian Liu\altaffilmark{3},
Lu Zhang\altaffilmark{4}, Wenyuan Cui\altaffilmark{1}, Yanchun Liang\altaffilmark{5},
Ping Niu\altaffilmark{1,6} and Bo Zhang\altaffilmark{1,*}}
%\and
%\author{\altaffilmark{}}
%\affil{}
%\email{} %% non-output

%% Alternate Affilations
\altaffiltext{1}{Department of Physics, Hebei Normal University,
NO. 20 Road East of 2nd Ring South, Shijiazhuang, 050024, China}
\altaffiltext{2}{School of Sciences, Hebei University of Science and Technology,
Shijiazhuang, 050018, China}
\altaffiltext{3}{Astronomy Department, Beijing Normal University, Beijing, 100875, China}
\altaffiltext{4}{College of Mathematics and Information Science, Hebei Normal University,
Shijiazhuang, 050024, China}
\altaffiltext{5}{Key Laboratory of Optical Astronomy, National Astronomical Observatories,
Chinese Academy of Sciences, Beijing, 100012, China}
\altaffiltext{6}{Department of Physics, Shijiazhuang University, Shijiazhuang 050035, China}
\altaffiltext{*}{E-mail address: zhangbo@mail.hebtu.edu.cn}


\begin{abstract}

The abundance scatter of heavier r-process elements ($Z \geqslant 56$)
relative to Fe ([r/Fe]) in metal-poor stars preserves excellent
information of the star formation history and provides important
insights into the various situations of the Galactic chemical
enrichment. In this respect, the upper and lower boundaries of
[r/Fe] could present useful clues for investigating the extreme
situations of the star formation history and the early Galactic
chemical evolution. In this paper, we investigate the formation of
the upper and lower boundaries of [r/Fe] for the gas clouds. We find
that, for a cloud from which metal-poor stars formed, the formation
of the upper limits of [r/Fe] is mainly due to the pollution from a
single main r-process event. For a cloud from which metal-poor stars
formed, the formation of the lower limits of [r/Fe] is mainly due to
the pollution from a single SN II event that ejects
primary Fe.

\end{abstract}





\section{Introduction}

The neutron-capture (n-capture) process which dominantly creates
heavy elements consists of the slow n-capture process (s-process) and
the rapid n-capture process (r-process) \citep{burbidge1957}. The
s-process consists of the weak s-process and the main s-process. Massive
stars are the astrophysical sites of the weak s-process
which mainly produces the lighter n-capture elements (e.g., Sr, Y
and Zr) \citep{lamb1977,raiteri1991,raiteri1993}. Asymptotic
giant branch stars are the astrophysical sites of the main
s-process which produces both the lighter and heavier n-capture
elements \citep{busso1999}. The r-process consists of the weak
r-process (or ``lighter element primary process'' (LEPP),
\citealt{travaglio2004}) and main r-process \citep{cowan1991}.
Because the yields of both the weak r-process (or LEPP) elements
and the light elements have primary nature, the weak r-process
is suggested to occur in SNe II with
the progenitor mass $M > 10 M_{\odot}$ \citep{travaglio2004,montes2007}.
The main r-process is considered to associate with the final stages of
massive star evolution, yet the actual astrophysical sites have
not been confirmed \citep{sneden2008,ishimaru2015,goriely2016}.
Two astrophysical sites of the main r-process are paid attention:
(1) the SNe II with the progenitor mass $M \approx 8-10 M_{\odot}$
\citep{travaglio1999,wanajo2003,qian2007}
and (2) neutron star mergers (NSMs)
\citep{lattimer1974,eichler1989,tsujimoto2014a,tsujimoto2014b}.
The r-process abundances of the solar system have been obtained
by \citet{kappeler1989} and \citet{arlandini1999} using the
residual approach. The two components of n-capture process have
been found in the elemental abundances of metal-poor stars
\citep{wasserburg1996,qian1998}. Based on the elemental abundances
of the main r-process stars and the weak r-process stars, adopting
the iterative method, \citet{li2013b} and \citet{hansen2014} derived
the abundances of the main r-process and the weak r-process.

Elemental abundances of metal-poor stars reflect the chemical
composition of the natal clouds polluted by different
nucleosynthetic processes. \citet{truran1981} studied the
abundances of the r-process elements in the Galactic halo stars
and suggested that the r-process enrichment in metal-poor stars
originates from the early massive stars. Because of the secondary
nature (i.e., the yields are correlated with the initial stellar
metallicity), the s-process contributions to the n-capture
elements in interstellar medium (ISM) are negligible at low
metallicities ([Fe/H] $\leq -1.5$). In
this case, the n-capture elements of metal-poor stars dominantly
come from the r-process \citep{travaglio1999,burris2000}. Study of the
abundance characteristics of the n-capture elements in metal-poor
stars is significant for understanding the r-process nucleosynthesis
and the chemical enrichment history of the early Galaxy.
\citet{gilroy1988} reported that there exists a large [r/Fe] scatter
for the heavier n-capture elements at low
metallicity. They proposed that the scatter reflects the
inhomogeneous mixing of products from different nucleosynthetic
processes. The large [r/Fe] scatter is also shown in more observational
work \citep[e.g.,][]{mcwilliam1995,ryan1996,mcwilliam1998,
sneden1998,burris2000,honda2004,barklem2005,francois2007,roederer2010}.
\citet{mathews1992} and \citet{travaglio1999} calculated the
Galactic evolution of the n-capture elements and suggested that the
abundances of the heavier r-process elements should be produced by
the lower-mass SNe II. \citet{sneden1994} studied the abundances of
the main r-process star CS 22892-052 and ascribed the Eu enrichment
in this star to the unmixed chemical composition of the cloud swept
up by the SN event. \citet{mcwilliam1995} and \citet{mcwilliam1998}
analyzed the chemical abundances of extremely metal-poor stars.
They proposed that the observed abundance
dispersions of the heavier n-capture elements indicate the chemical
inhomogeneities of the ISM. However, \citet{ryan1996} suggested
that the different gas clouds would be entirely inhomogeneous and
the r-process element enrichment in a gas cloud is only due
to the pollution from a main r-process event. \citet{tsujimoto1999}
studied the [Eu/Fe] scatter in the Galactic halo and suggested
that the [Eu/Fe] scatter in metal-poor stars is the integrated
results of remnants of the SNe with different progenitor mass.
\citet{argast2000} analyzed the [Eu/Fe] pattern of metal-poor stars using
the three-dimensional stochastic evolution model. They proposed that
(1) the large [Eu/Fe] scatter for [Fe/H] $< -3.0$ should result from
the inhomogeneities of the clouds swept up by the
SNe II with different initial mass and (2) the [Eu/Fe] scatter
decreases with increasing metallicity because of the chemical mixing of
the ISM. \citet{travaglio2001} calculated the Galactic halo evolution
considering the chemical mixing of the clouds
and the products of SNe. They found that the [r/Fe] scatter at low
metallicity is the results of inhomogeneous mixing of the ISM and
the chemical enrichment of the gas clouds is dominated by the star
formation episodes. \citet{fields2002} presented a simple model to
explain the [Eu/Fe] scatter for metal-poor stars and suggested that
the decreasing [Eu/Fe] scatter with increasing metallicity is due to
the mixing of the sources produced by different nucleosynthetic events.
\citet{cescutti2008} used the stochastic evolution model to
investigate the formation of the [r/Fe] scatter in metal-poor stars
and suggested that the different [r/Fe] dispersions are caused by
the massive stars with different mass ($12 \sim 30 M_{\odot}$).

Over the years, NSM is also argued as a plausible astrophysical site
of main r-process which leads to the r-process enrichment in
metal-poor stars. Based on the observed abundances of lighter
r-process elements (i.e., Y and Zr) and heavier r-process element Eu
of metal-poor stars, \citet{tsujimoto2014a} suggested that (1) the
core-collapse supernovae produce the lighter r-process
elements and (2) there exist two types of NSMs: one type only
produces the heavier r-process elements and the other produces both
the lighter and heavier r-process elements. Considering the
accretion of ISM and the chemical enrichment of intergalactic medium,
\citet{komiya2014} calculated the chemical evolution of the
heavier r-process elements Ba and Eu. They found that both the SNe
II and NSMs can reproduce the abundance patterns of the heavier
r-process elements of extremely metal-poor stars.
\citet{tsujimoto2014b} studied the Eu enrichment in the Galactic
halo and proposed that the [Eu/Fe] scatter in the Galactic halo
stars can be explained by the hierarchical galaxy formation model.
Furthermore, they suggested that NSMs should be the dominant
astrophysical site of the main r-process. Through simulating the
chemical and dynamic evolution of NSMs, \citet{rosswog2014}
suggested that the dynamic ejecta of NSMs could produce the heavier
r-process elements with $A > 130$.
In order to distinguish the products of the SN II and
NSM events, \citet{hotokezaka2015} calculated the abundances of the
short-lived element $^{244}$Pu and compared the theoretical values
with the observed abundances. They proposed that the NSM events can
naturally explain the $^{244}$Pu abundances and are an excellent
candidate of the astrophysical origins of the heavier r-process
elements. By comparing the calculated abundances with the observed
abundances of CS 22892-052, \citet{ramirez2015} concluded that
NSMs are a favorable source of the heavier r-process elements in
CEMP-r stars. They also found that the NSM events could naturally
explain the scatter of observed
Eu abundances in CEMP-r stars. Based on the discovery of strong
r-process enhanced stars in the ultra-faint dwarf galaxy Reticulum
II, \citet{ji2016} suggested that the r-process material in
Reticulum II should be produced by the single NSM
event. For examining the NSMs as astrophysical origins of the
r-process elements in dwarf galaxies, \citet{beniamini2016a}
calculated the proper motion and the time until merger of neutron
star binaries. They concluded that (1) more than 50$\%$ of neutron
star binaries have small proper motion to avoid the pollution of SN
ejecta and (2) more than 90$\%$ of neutron star binaries merge
within 300 Myr. These results indicate that NSMs could be
naturally responsible for the observed r-process abundances of the
metal-poor stars of the early Galaxy. Based on the hierarchical
chemical evolution model, \citet{komiya2016} simulated
the abundances of the r-process elements for metal-poor stars after
considering the pollution of NSM ejecta. They suggested that the NSM
scenario can successfully reproduce the [r/Fe] scatter for
metal-poor stars. Through calculating the rate and yields for the
r-process event in dwarf galaxies, \citet{beniamini2016b} found
that the dwarf galaxies and the Milky Way share the same r-process
mechanism.

For metal-poor stars, the $\alpha$ elements (e.g., Mg, Si, Ca and
Ti) and Fe originate from massive stars
\citep{woosley1995,kobayashi2006, heger2010,mishenina2013}. To
explore the abundance correlations between $\alpha$ elements and
main r-process elements, \citet{li2014} plotted the observed
[$\alpha$/Eu] as a function of [Eu/Fe] for metal-poor stars and
found that the observed [$\alpha$/Eu] ratios decrease linearly with
increasing [Eu/Fe] and the slope is close to $-1$. The results mean
that the abundances of $\alpha$ elements and Fe have no correlation
with those of Eu, which indicates that the astrophysical site
producing main r-process elements does not produce $\alpha$ elements
and Fe. The observational and theoretical results of the previous
work imply that the [r/Fe] scatter is dominated by more than one
factor. Obviously, the upper and lower boundaries of [r/Fe] could
provide useful clues about the extreme situations of the chemical
enrichment of the gas clouds. So the quantitative study of the
[r/Fe] boundaries in metal-poor stars is important. In this paper,
we investigate the formation of the [r/Fe] boundaries for the
heavier n-capture elements ($Z \geq 56$) of metal-poor stars in
\S\,2. Conclusions are presented in \S\,3.

\section{Abundance Boundaries of [r/Fe] for the Heavier n-capture
Elements in Metal-poor Stars}

\begin{figure*}
\centering
%\vspace{-91cm}
\includegraphics[width=16cm]{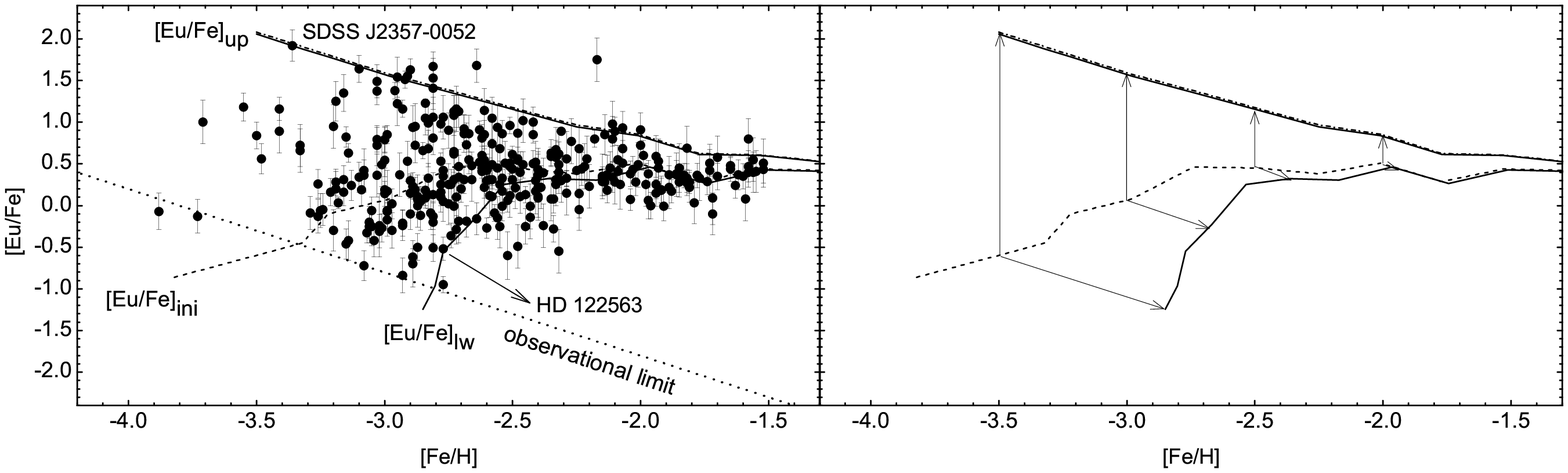}
\vspace{-21cm}
\caption {[Eu/Fe] ratios as a function of [Fe/H] for
the metal-poor stars. In the left panel, the filled circles
represent the observed abundance ratios. The dashed line refers
to the initial [Eu/Fe] ratios of the clouds. The upper solid line
and dotted-dashed line are the calculated upper boundaries of
[Eu/Fe] for the clouds swept up by the SNe II with the progenitor
mass $M \approx 8-10 M_{\odot}$ and NSMs, respectively.
The lower solid line is the calculated lower boundary of
[Eu/Fe]. The dotted line with $-1$ slope represents the
sensitivity limit for observing Eu. In the right panel, the
lines have the same meaning as in the left
panel. Each up arrow represents the jump of [Eu/Fe] in a
cloud swept up by a single main r-process event. Each inclined
arrow represents the jumps of [Eu/Fe] and [Fe/H] in a
cloud swept up by a single SN II-Fe event.}
\label{f1.eps}
\end{figure*}

Generally, Eu is deemed as a typical main r-process element in the
solar system. The left panel of Figure \ref{f1.eps} shows the
observed [Eu/Fe] as a function of [Fe/H] for the metal-poor stars
\citep{westin2000,cowan2002,hill2002,johnson2002,sneden2003,christlieb2004,
honda2004,barklem2005,ivans2006,honda2006,honda2007,francois2007,lai2008,hayek2009,
aoki2010,mashonkina2010,roederer2010,roederer2014,hollek2011,hansen2012,hansen2015,
siqueira2012,cohen2013,yong2013,mashonkina2014,jacobson2015,li2015a,li2015b,siqueira2016}.
In order to avoid the effect of s-process, we select the sample
stars with [Fe/H] $\leqslant -1.5$ and [Ba/Eu] $< 0$. The dotted line
with $-1$ slope in the panel represents the sensitivity limit for
the observing Eu, which is 0.8 dex lower than that adopted by
\citet{travaglio2001}. From the panel we can see that the [Eu/Fe]
scatter is larger than 2 dex at [Fe/H] $= -3.0$, whereas the scatter
decreases with increasing metallicity. The observed upper boundary
of [Eu/Fe] is explicit: it is close to a straight line with the
slope $\thicksim -1$ for [Fe/H] $\lesssim -2.5$, while the
decreasing trend flattens for [Fe/H] $> -2.5$. On the other hand,
the observed lower boundary shows a sharp increasing trend for
[Fe/H] $\lesssim -2.5$, while the increasing trend flattens for
[Fe/H] $> -2.5$.

In order to explain the upper and lower boundaries of [Eu/Fe] for
the metal-poor stars, we consider the initial chemical composition
of the gas clouds and the pollution from the two nucleosynthetic
events: (1) the main r-process event and (2) the SN II event that
ejects primary Fe (hereafter simply the SN II-Fe event). The main
r-process event (e.g., the SN II with the progenitor mass $M \approx
8-10 M_{\odot}$ or NSM) produces and ejects the main r-process
elements \citep{travaglio1999,qian2007,tsujimoto2014b}. Whereas
the SN II-Fe event (i.e., the SN II with the progenitor mass
$M > 10 M_{\odot}$) mainly produces the weak r-process (or LEPP)
elements and ejects light elements and Fe group elements
simultaneously \citep{travaglio2004,montes2007,li2013a}.

For the main r-process events, although two possible sites
(i.e., the SNe II with the progenitor mass $M \approx 8-10 M_{\odot}$
or NSMs) attract common attention, the actual main r-process
sites and corresponding main r-process yields have not been determined.
For a gas cloud swept up by a single main r-process event,
the relationship between the Eu abundance $N_{Eu}$ and the Fe
abundance $N_{Fe}$ is
%----------------------------------------------------------------
\begin{equation}
%\label{eq1}
\frac{N_{Eu}}{N_{Fe}}=(\frac{A_{Fe}}{A_{Eu}})(\frac{M_{SW,rm}X_{Eu}+Y_{Eu}}{M_{SW,rm}X_{Fe}}),
\end{equation}
%----------------------------------------------------------------
where $M_{SW,rm}$ is the mass of the cloud. $X_{Eu}$ and $X_{Fe}$
are the initial mass fractions of Eu and Fe in the cloud. $Y_{Eu}$
is the Eu yield of the main r-process event. $A_{Eu}$ and $A_{Fe}$
are the atomic weights of Eu and Fe, respectively. Based on the
observed abundances of Ba and Eu, \citet{cescutti2006} computed the
mean [Ba/Fe] and [Eu/Fe] ratios in different [Fe/H] bins. In this
work, we adopt the mean [Eu/Fe] ratios presented by
\citet{cescutti2006} as the initial abundance ratios ([Eu/Fe]$_{ini}$)
of the gas clouds, which are plotted in the left panel of
Figure \ref{f1.eps} by the dashed line. The upper limits of [Eu/Fe]
of the clouds are calculated for two cases.

Case A: the pollution from the SNe II with the progenitor mass $M
\approx 8-10 M_{\odot}$. The core-collapse SNe should consist of two
routes \citep[e.g.,][]{qian2007}: (1) the Fe core-collapse SNe with
the progenitor mass $M \gtrsim 11 M_{\odot}$ from which the light
elements and Fe group elements are ejected and (2) the O-Ne-Mg
core-collapse SNe with the progenitor mass $M \approx 8-10
M_{\odot}$ in which the light elements and Fe group elements are not
produced. The main r-process may take place in O-Ne-Mg core-collapse
SNe. In this case, the astrophysical sites ejecting main r-process
elements do not eject light elements and Fe group elements. Owing to
the high Eu abundance ([Eu/Fe] $= 1.92$) and low metallicity ([Fe/H] $=
-3.36$), the sample star SDSS J2357-0052 \citep{aoki2010} is taken as
the representative star and [Eu/Fe] $= 1.92$ is taken as the upper
limit at [Fe/H] $= -3.36$. Adopting the Eu yield of the main
r-process event $Y_{Eu} = 6.4\times10^{-7} M_{\odot}$
\citep{travaglio2001}, from equation (1) we can derive $M_{SW,rm} =
4.5 \times 10^{4} M_{\odot}$. Using the derived cloud mass, from
equation (1) we can derive the upper limits ([Eu/Fe]$_{up}$) for the
clouds with different metallicities which were swept up by the SNe
II with the progenitor mass $M \approx 8-10 M_{\odot}$. The
calculated results are shown in the left panel of Figure
\ref{f1.eps} by the solid line. We can see that the calculated upper
boundary is close to a straight line with the slope $\thicksim -1$
for [Fe/H] $\lesssim -2.5$, while the decreasing trend flattens for
[Fe/H] $> -2.5$. Obviously, the calculated results are consistent
with the observed upper boundary of [Eu/Fe] of the sample stars.

Case B: the pollution from NSMs. Based on the NSM scenario,
\citet{komiya2016} explored the [r/Fe] scatter of metal-poor stars
using the hierarchical galaxy formation model. They suggested that
the Eu yield of a NSM event is about $1.5\times10^{-4} M_{\odot}$
and the mass of the cloud swept up by the single main r-process
event is about $10^{7} M_{\odot}$. For the NSM scenario, the Eu
yield is about 3 orders of magnitude higher than what is adopted for
Case A and the gas mass that Eu is diluted into is also about 3
orders of magnitude higher than what is adopted for Case A. Adopting
the Eu yield $Y_{Eu} = 1.5\times10^{-4} M_{\odot}$ and the polluted
cloud mass $M_{SW,rm} = 10^{7} M_{\odot}$, we calculate the upper
limits of [Eu/Fe] for the clouds swept up by NSMs. The calculated
results are plotted in the left panel of Figure \ref{f1.eps} by the
dotted-dashed line. We can see that the upper boundaries of [Eu/Fe] for
the two cases are very similar.

For illustrating the formation of the upper boundary of [Eu/Fe]
clearly, in the right panel we use the up arrows to show the jumps
of [Eu/Fe]. For the cloud with the initial ratios [Eu/Fe]$_{ini} = -0.59$
and [Fe/H] $= -3.5$, the final ratio [Eu/Fe]$_{up} \approx 2.06$ means
that the [Eu/Fe] ratio jumps from $-0.59$ to 2.06 because of the
pollution from the single main r-process event. In this case, the
increasing value of [Eu/Fe] reaches about 2.65 dex, since (1) the initial Fe
mass fraction $X_{Fe}$ of the cloud is small and (2) the initial Eu
mass in the cloud is much lower than the Eu yield of the main
r-process event (i.e., $M_{SW,rm}X_{Eu} \ll Y_{Eu}$).
On the other hand, for the cloud with the initial ratios
[Eu/Fe]$_{ini} = 0.44$ and [Fe/H] $= -1.5$, the final ratio
[Eu/Fe]$_{up} \approx 0.60$ means that the increasing value of [Eu/Fe] is
only about 0.16 dex. The two reasons lead to the flat of the upper boundary
of [Eu/Fe]: (1) the cloud contains more Fe and (2) the initial Eu mass
in the cloud is higher than the Eu yield of the main r-process event
(i.e., $M_{SW,rm}X_{Eu} > Y_{Eu}$). The derived results mean that,
for a cloud from which metal-poor stars formed, the formation of the
upper limit of [Eu/Fe] is mainly due to the pollution from a single
main r-process event.

For a gas cloud swept up by a single SN II-Fe event, the
relationship between the Eu abundance $N_{Eu}$ and the Fe abundance
$N_{Fe}$ is
%----------------------------------------------------------------
\begin{equation}
%\label{eq2}
\frac{N_{Eu}}{N_{Fe}}=(\frac{A_{Fe}}{A_{Eu}})(\frac{M_{SW,pri}X_{Eu}}{M_{SW,pri}X_{Fe}+Y_{Fe}}),
\end{equation}
%----------------------------------------------------------------
where $M_{SW,pri}$ is the mass of the cloud. $Y_{Fe}$ is the Fe
yield of the SN II-Fe event. We take the typical weak r-process star
HD 122563 ([Fe/H] $= -2.77$, [Eu/Fe] $= -0.52$, \citealt{honda2006})
as the representative star and take [Eu/Fe] $= -0.52$ as the lower
limit at [Fe/H] $= -2.77$. Adopting the polluted cloud mass
$M_{SW,pri} = 6.5\times10^{4} M_{\odot}$ \citep{tsujimoto1999},
from equation (2) we
can derive $Y_{Fe} = 0.1 M_{\odot}$. Using the derived Fe yield,
from equation (2) we can derive the lower limits ([Eu/Fe]$_{lw}$) for
the clouds with different metallicities, which are shown in the left
panel of Figure \ref{f1.eps} by the solid line. We can see that the
calculated lower boundary shows a sharp increasing trend for [Fe/H]
$\lesssim -2.5$ and a mild increasing trend for [Fe/H] $> -2.5$.
There are some observed [Eu/Fe] ratios that are significantly lower
than the calculated lower boundary in the range $-2.5 <$ [Fe/H] $<
-2.2$. Note that the initial [Eu/Fe] ratio of a cloud takes some
effect on the lower limit of [Eu/Fe]. In this work, we take the
observed mean [Eu/Fe] ratios as the initial [Eu/Fe] ratios of the
clouds. Obviously, the actual initial ratios of the clouds could
deviate from the mean ratios. If the actual initial [Eu/Fe] ratio
of a cloud swept up by a SN II-Fe event is lower than the mean
[Eu/Fe] ratio, the actual lower limit of [Eu/Fe] should be lower
than the calculated lower limit plotted in Figure \ref{f1.eps}.
The observed low [Eu/Fe] ratios should imply that the actual initial
[Eu/Fe] ratios of these clouds are lower than the mean [Eu/Fe] ratios.
Furthermore, because the explosion energy of SNe II depends on their
progenitor mass \citep{kobayashi2006,nomoto2013} and the mass of the
clouds swept up by the SNe II increases with increasing explosion
energy \citep{machida2005}, the mass of the clouds swept up by the
SNe II with the progenitor mass $M > 10 M_{\odot}$ should be larger
than that of the clouds swept up by the SNe II with the progenitor
mass $M \approx 8-10 M_{\odot}$.

For illustrating the formation of the lower boundary of [Eu/Fe]
clearly, in the right panel we use the inclined arrows to represent
the jumps of [Eu/Fe] and [Fe/H]. For the cloud with the initial
ratios [Eu/Fe]$_{ini} = -0.60$ and [Fe/H] $= -3.5$, the final
ratios [Eu/Fe]$_{lw} = -1.25$ and [Fe/H] $= -2.85$ imply that the
[Fe/H] ratio jumps from $-3.5$ to $-2.85$ because of the pollution
from the single SN II-Fe event. In this case, the increasing value
of [Fe/H] reaches 0.65 dex, since the initial Fe mass in the cloud
is lower than the Fe yield of the SN II-Fe event (i.e.,
$M_{SW,pri}X_{Fe} < Y_{Fe}$). On the other hand, for the cloud with
initial ratios [Eu/Fe]$_{ini} = 0.44$ and [Fe/H] $= -1.5$, the
final ratios [Eu/Fe]$_{lw} = 0.42$ and [Fe/H] $= -1.48$ mean that
the increasing value of [Fe/H] is only 0.02 dex. The two reasons
lead to the flat of the lower boundary of [Eu/Fe]: (1) the cloud
contains more Eu and (2) the initial Fe mass in the cloud is higher
than the Fe yield of the SN II-Fe event (i.e., $M_{SW,pri}X_{Fe} >
Y_{Fe}$). The derived results mean that, for a cloud from which
metal-poor stars formed, the formation of the lower limit of [Eu/Fe]
is mainly due to the pollution from a single SN II-Fe event.

The first generation of very massive stars only produces light
elements and Fe group elements \citep{komiya2014}. This abundance
pattern is represented by the prompt (P) component \citep{qian2001}
which should be responsible for the origin of light elements and Fe
for [Fe/H] $\lesssim -3.5$. For the clouds with [Fe/H] $\gtrsim -3.5$,
the element Eu originates from the main r-process events (e.g., the
SNe II with the progenitor mass $M \approx 8-10 M_{\odot}$ or NSMs)
and the element Fe originates from (1) the prompt inventory
(P-inventory) of Fe at the early universe \citep{qian2001} and (2)
the SN II-Fe events. The effect of the P-inventory became smaller
when the SN II-Fe events began to pollute the ISM \citep{qian2001,li2013b}.
On the other hand, for the most stars with [Fe/H] $\lesssim -3.5$,
the element Fe should mainly originate from the P-inventory
and the observed r-process abundances should be ascribed to the
surface pollution by the ISM which had contained r-process material
\citep{komiya2014,komiya2016}.

\begin{figure*}
\centering
\includegraphics[width=13cm]{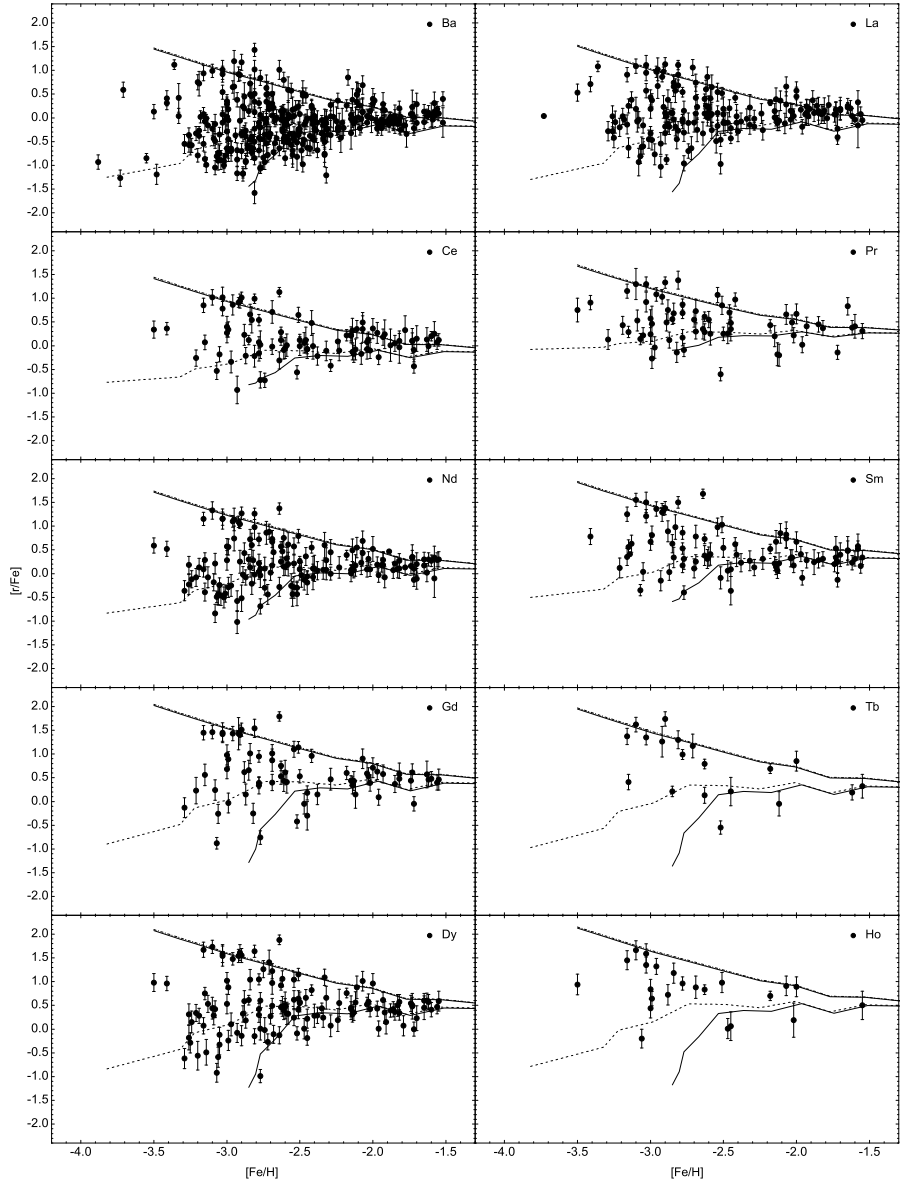}
\vspace{-4cm}
\caption {The same as in the left panel of Figure \ref{f1.eps}, but for the
[r/Fe] ratios (r: Ba, La, Ce, Pr, Nd, Sm, Gd, Tb, Dy and Ho).}
\label{f2.eps}
\end{figure*}

\begin{figure*}
\centering
%\vspace{-1cm}
\includegraphics[width=13cm]{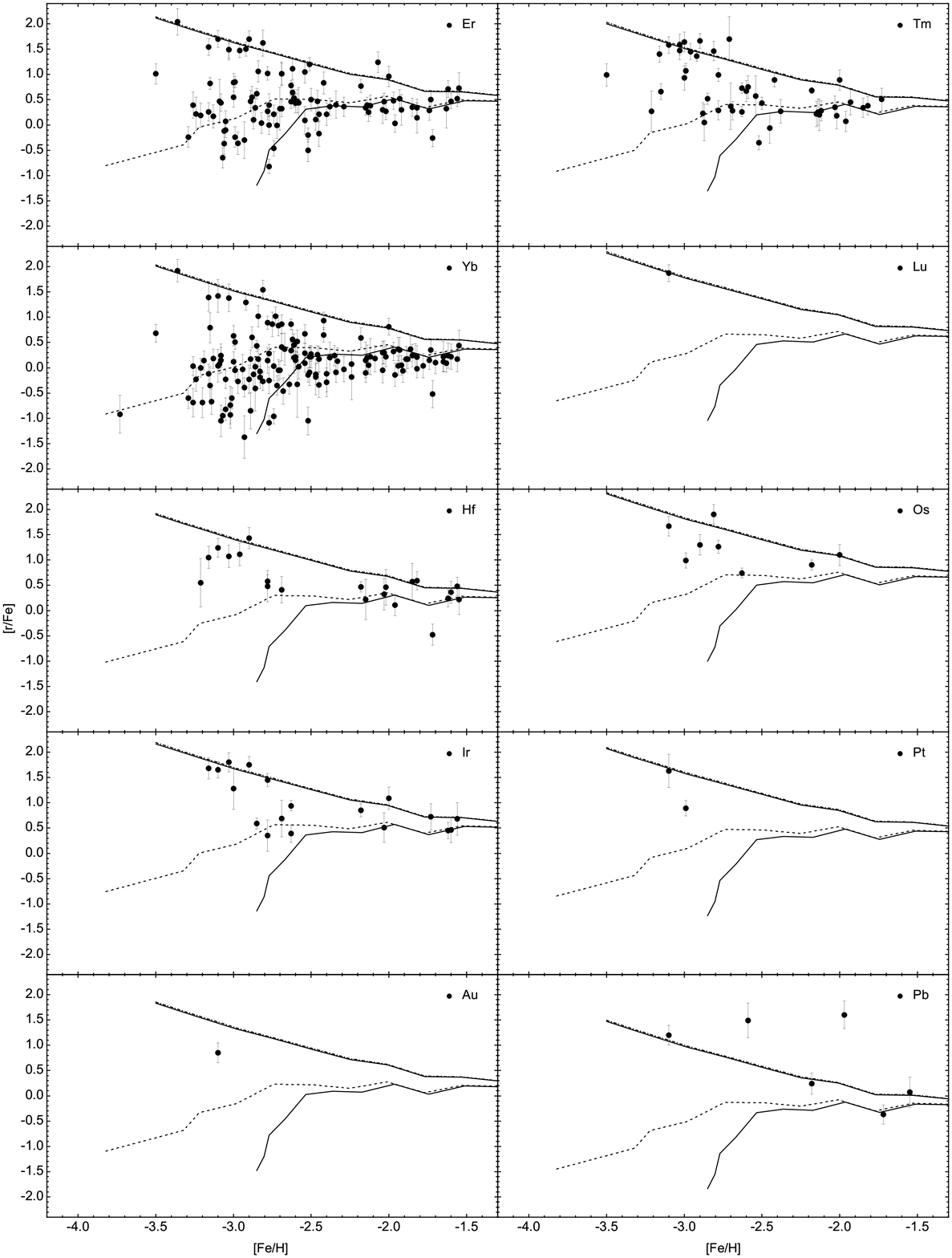}
\vspace{-4cm}
\caption {The same as in the left panel of Figure \ref{f1.eps}, but for the
[r/Fe] ratios (r: Er, Tm, Yb, Lu, Hf, Os, Ir, Pt, Au and Pb).}
\label{f3.eps}
\end{figure*}

Figures \ref{f2.eps} and \ref{f3.eps} show the observed [r/Fe] (r:
Ba, La, Ce, Pr, Nd, Sm, Gd, Tb, Dy, Ho, Er, Tm, Yb, Lu, Hf, Os, Ir,
Pt, Au and Pb) as a function of [Fe/H] for the metal-poor stars.
From the figures we can see that the observed trends of the upper
and lower boundaries of [r/Fe] are similar to those of [Eu/Fe]. For
a gas cloud swept up by a single main r-process event, the
relationship between the r-process abundance $N_{i,r}$ and the Fe
abundance $N_{Fe}$ is:
%----------------------------------------------------------------
\begin{equation}
%\label{eq3}
\frac{N_{i,r}}{N_{Fe}}=(\frac{A_{Fe}}{A_{i}})(\frac{M_{SW,rm}X_{i}+Y_{i,rm}}{M_{SW,rm}X_{Fe}}),
\end{equation}
%----------------------------------------------------------------
where $X_{i}$ is the initial mass fraction of the $i$th element.
$Y_{i,rm}$ is the main r-process yield. $A_{i}$ is the atomic weight.
The main r-process yield $Y_{i,rm}$ can be derived as:
%----------------------------------------------------------------
\begin{equation}
%\label{eq4}
Y_{i,rm}=\frac{A_{i}N^{*}_{i,rm}}{A_{Eu}N^{*}_{Eu}}Y_{Eu},
\end{equation}
%----------------------------------------------------------------
where $N^{*}_{i,rm}$ and $N^{*}_{Eu}$ are the component abundances,
which are adopted from \citet{li2013b}. Using the initial [Eu/Fe]
ratios of the clouds adopted from \citet{cescutti2006}
and the component abundances adopted from \citet{li2013b}, we can
derive the initial [r/Fe] ratios for the clouds, which are
plotted in Figures \ref{f2.eps} and \ref{f3.eps} by the dashed lines.
The upper limits of [r/Fe] of the clouds are also calculated for two
cases.

Case A: the pollution from the SNe II with the progenitor mass
$M \approx 8-10 M_{\odot}$.
Adopting the Eu yield $Y_{Eu} = 6.4\times10^{-7} M_{\odot}$ and the
polluted cloud mass $M_{SW,r,m} = 4.5 \times 10^{4} M_{\odot}$, from
equations (3) and (4) we can derive the upper limits of [r/Fe]
for the clouds swept up by the SNe II with the progenitor mass
$M \approx 8-10 M_{\odot}$. The results are plotted in Figures
\ref{f2.eps} and \ref{f3.eps} by the solid lines.

Case B: the pollution from NSMs. Adopting the Eu yield $Y_{Eu} =
1.5\times10^{-4} M_{\odot}$ and the polluted cloud mass
$M_{SW,rm} = 10^{7} M_{\odot}$, we derive the upper limits
of [r/Fe] for the clouds swept up by NSMs, which are plotted
in Figures \ref{f2.eps} and \ref{f3.eps} by the dotted-dashed lines.
Obviously, the calculated upper boundaries are close to the observed
upper boundaries of [r/Fe] of the sample stars. These results could
provide more supports for the suggestion that the formation of the
upper limits of [r/Fe] in a cloud is mainly due to the pollution
from a single main r-process event.

For a gas cloud swept up by a single SN II-Fe event, the
relationship between the r-process abundance $N_{i,r}$ and the Fe
abundance $N_{Fe}$ is:
%----------------------------------------------------------------
\begin{equation}
%\label{eq2}
\frac{N_{i,r}}{N_{Fe}}=(\frac{A_{Fe}}{A_{i}})(\frac{M_{SW,pri}X_{i}+Y_{i,pri}}{M_{SW,pri}X_{Fe}+Y_{Fe}}),
\end{equation}
%----------------------------------------------------------------
where $Y_{i,pri}$ is the primary yield of the $i$th
element from the SN II-Fe event. The primary yield $Y_{i,pri}$
can be derived as:
%----------------------------------------------------------------
\begin{equation}
%\label{eq4}
Y_{i,pri}=\frac{A_{i}N^{*}_{i,pri}}{A_{Fe}N^{*}_{Fe}}Y_{Fe},
\end{equation}
%----------------------------------------------------------------
where $N^{*}_{i,pri}$ and $N^{*}_{Fe}$ are the component abundances,
which are adopted from \citet{li2013b}. Adopting the derived Fe yield
$Y_{Fe} = 0.1 M_{\odot}$ and the polluted cloud mass
$M_{SW,pri} = 6.5\times10^{4} M_{\odot}$, from equations (5) and (6)
we can derive the lower limits of [r/Fe] for the clouds with
different metallicities. The results are plotted in Figures
\ref{f2.eps} and \ref{f3.eps} by the solid lines. We can see that
the lines are close to the observed lower boundaries of [r/Fe] of
the sample stars. The results also mean that the formation of the
lower limits of [r/Fe] in a cloud is mainly due to the pollution
from a single SN II-Fe event.

\section{Conclusions}

The [r/Fe] scatter of the heavier n-capture elements in metal-poor
stars preserves excellent information of the star formation history
and provides important insights into the various situations of the
Galactic chemical enrichment. In this respect, the upper and lower
boundaries of [r/Fe] could present useful clues for investigating
the extreme situations of the chemical enrichment of the clouds. In
this paper, we investigate the formation of the upper and lower
boundaries of [r/Fe] for the clouds. The main results are listed as
follows:

1. Based on the assumptions of (1) the progenitor of the main
r-process event does not produce Fe and (2) the yields of the main
r-process event possess the primary nature (i.e., the yields are
uncorrelated with the initial stellar metallicity), we find that the
observed upper boundaries of [r/Fe] can be explained. For the clouds
with the initial metallicities [Fe/H] $\lesssim -2.5$, the upper
boundaries of [r/Fe] are close to straight lines with slopes
$\thicksim -1$, since the mass of the initial r-process elements in
each of the clouds is lower than the yields of the single main
r-process event. On the other hand, for the clouds with the initial
metallicities [Fe/H] $> -2.5$, the upper boundaries of [r/Fe] show
mild decreasing trends, since the mass of the initial r-process
elements in each of the clouds is close to or higher than the yields
of the single main r-process event.

2. Based on the assumptions of (1) the progenitor of the SN II-Fe
event does not produce the main r-process elements and (2) the yields of
the SN II-Fe event possess the primary nature, we find that the
observed lower boundaries of [r/Fe] can be explained. For the clouds
with the initial metallicities [Fe/H] $\lesssim -2.5$, the lower
boundaries of [r/Fe] show sharp increasing trends, since the initial
Fe mass in each of the clouds is lower than the Fe yield of the
single SN II-Fe event. On the other hand, for the clouds with the
initial metallicities [Fe/H] $> -2.5$, the lower boundaries of [r/Fe]
show mild increasing trends, since the initial Fe mass in each of
the clouds is close to or higher than the Fe yield of the single SN
II-Fe event.

3. The observed upper and lower boundaries of [r/Fe] present the
extreme situations of the chemical enrichment in the early Galaxy.
The calculated results mean that, for a cloud from which
metal-poor stars formed, the formation of the upper limits of [r/Fe]
is mainly due to the pollution from a single main r-process event.
For a cloud from which metal-poor stars formed, the formation of
the lower limits of [r/Fe] is mainly due to the pollution from a
single SN II-Fe event.

Our results may provide useful clues for investigating the star
formation history and the early Galactic chemical evolution.
Obviously, more observational and theoretical studies of the main
r-process event and the SN II-Fe event are desirable.

%% Included in this acknowledgments section are examples of the
%% AASTeX hypertext markup commands. Use \url without the optional [HREF]
%% argument when you want to print the url directly in the text. Otherwise,
%% use either \url or \anchor, with the HREF as the first argument and the
%% text to be printed in the second.

\acknowledgments

This work has been supported by the National Natural Science
Foundation of China under Grants 11673007, 11273011, U1231119,
10973006, 11403007, 11547041, 11643007 and 11273026, the Natural
Science Foundation of Hebei Province under Grant A2011205102, the
Program for Excellent Innovative Talents in University of Hebei
Province under Grant CPRC034, and the Innovation Fund Designated
for Graduate Students of Hebei Province under Grant sj2016023.

%% in parentheses, after the keyword, but they will not be verified.



%---------------------------------------------------------------
%---------------------------------------------------------------------------------
%---------------------------------------------------------------------------------
%---------------------------------------------------------------------------------


\begin{thebibliography}{}
\bibitem[\protect\citeauthoryear{Aoki et al.}{2010}]{aoki2010} Aoki, W., Beers, T. C., Honda, S., et al. 2010, ApJL, 723, L201
\bibitem[\protect\citeauthoryear{Argast et al.}{2000}]{argast2000} Argast, D., Samland, M., Gerhard, O. E., \& Thielemann, F. K. 2000, A\&A, 356, 873
\bibitem[\protect\citeauthoryear{Arlandini et al.}{1999}]{arlandini1999} Arlandini, C., K\"{a}ppeler, F., Wisshak, K., et al. 1999, ApJ, 525, 886
\bibitem[\protect\citeauthoryear{Barklem et al.}{2005}]{barklem2005} Barklem, P. S., Christlieb, N., Beers, T. C., et al. 2005, A\&A, 439, 129
\bibitem[\protect\citeauthoryear{Beniamini et al.}{2016a}]{beniamini2016a} Beniamini, P., Hotokezaka, K., \& Piran, T. 2016a, ApJL, 829, L13
\bibitem[\protect\citeauthoryear{Beniamini et al.}{2016b}]{beniamini2016b} Beniamini, P., Hotokezaka, K., \& Piran, T. 2016b, ApJ, 832, 149
\bibitem[\protect\citeauthoryear{Burbidge et al.}{1957}]{burbidge1957} Burbidge, E. M., Burbidge, G. R., Fowler, W. A., et al. 1957, RvMP, 29, 547
\bibitem[\protect\citeauthoryear{Burris et al.}{2000}]{burris2000} Burris, D. L., Pilachowski, C. A., Armandroff, T. E., et al. 2000, ApJ, 544, 302
\bibitem[\protect\citeauthoryear{Busso et al.}{1999}]{busso1999} Busso, M., Gallino, R., \& Wasserburg, G. J. 1999, ARA\&A, 37, 239
\bibitem[\protect\citeauthoryear{Christlieb et al.}{2004}]{christlieb2004} Christlieb, N., Beers, T. C., Barklem, P. S., et al. 2004, A\&A, 428, 1027
\bibitem[\protect\citeauthoryear{Cohen et al.}{2013}]{cohen2013} Cohen, J. G., Christlieb, N., Thompson, I., et al. 2013, ApJ, 778, 56
\bibitem[\protect\citeauthoryear{Cowan et al.}{1991}]{cowan1991} Cowan, J. J., Thielemann, F. K., \& Truran, J. W. 1991, PhR, 208, 267
\bibitem[\protect\citeauthoryear{Cowan et al.}{2002}]{cowan2002} Cowan, J. J., Sneden, C., Burles, S., et al. 2002, ApJ, 572, 861
\bibitem[\protect\citeauthoryear{Cescutti et al.}{2006}]{cescutti2006} Cescutti, G., Franc\c{o}is, P., Matteucci, F., et al. 2006, A\&A, 448, 557
\bibitem[\protect\citeauthoryear{Cescutti}{2008}]{cescutti2008} Cescutti, G. 2008, A\&A, 481, 691
\bibitem[\protect\citeauthoryear{Eichler et al.}{1989}]{eichler1989} Eichler, D., Livio, M., Piran, T., \& Schramm, D. N. 1989, Natur, 340, 126
\bibitem[\protect\citeauthoryear{Fields et al.}{2002}]{fields2002} Fields, B. D., Truran, J. W., \& Cowan, J. J. 2002, ApJ, 575, 845
\bibitem[\protect\citeauthoryear{Franc\c{o}is et al.}{2007}]{francois2007} Franc\c{o}is, P., Depagne, E., Hill, V., et al. 2007, A\&A, 476,935
\bibitem[\protect\citeauthoryear{Gilroy et al.}{1988}]{gilroy1988} Gilroy, K. K., Sneden, C., Pilachowski, C. A., \& Cowan, J. J. 1988, ApJ, 327, 298
\bibitem[\protect\citeauthoryear{Goriely \& Janka}{2016}]{goriely2016} Goriely, S., \& Janka H. T. 2016, MNRAS, 459, 4174G
\bibitem[\protect\citeauthoryear{Hansen et al.}{2014}]{hansen2014} Hansen, C. J., Montes, F., \& Arcones, A. 2014, ApJ, 797, 123H
\bibitem[\protect\citeauthoryear{Hansen et al.}{2012}]{hansen2012} Hansen, C. J., Primas, F., Hartman, H., et al. 2012, A\&A, 545, A31
\bibitem[\protect\citeauthoryear{Hansen et al.}{2015}]{hansen2015} Hansen, T., Hansen, C.J., Christlieb, N., et al. 2015, ApJ, 807, 173
\bibitem[\protect\citeauthoryear{Hayek et al.}{2009}]{hayek2009} Hayek, W., Wiesendahl, U., Christlieb, N., et al. 2009, A\&A, 504, 511
\bibitem[\protect\citeauthoryear{Heger \& Woosley}{2010}]{heger2010} Heger, A., \& Woosley, S. E. 2010, ApJ, 724, 341
\bibitem[\protect\citeauthoryear{Hill et al.}{2002}]{hill2002} Hill, V., Plez, B., Cayrel, R., et al. 2002, A\&A, 387, 560
\bibitem[\protect\citeauthoryear{Hollek et al.}{2011}]{hollek2011} Hollek, J. K., Frebel, A., Roederer, I. U., et al. 2011, ApJ, 742, 54
\bibitem[\protect\citeauthoryear{Honda et al.}{2004}]{honda2004} Honda, S., Aoki, W., Kajino, T., et al. 2004, ApJ, 607, 474
\bibitem[\protect\citeauthoryear{Honda et al.}{2006}]{honda2006} Honda, S., Aoki, W., Ishimaru, Y., et al. 2006, ApJ, 643, 1180
\bibitem[\protect\citeauthoryear{Honda et al.}{2007}]{honda2007} Honda, S., Aoki, W., Ishimaru, Y., \& Wanajo, S. 2007, ApJ, 666, 1189
\bibitem[\protect\citeauthoryear{Hotokezaka et al.}{2015}]{hotokezaka2015} Hotokezaka, K., Piran, T., \& Paul, M. 2015, NatPh, 11, 1042
\bibitem[\protect\citeauthoryear{Ishimaru et al.}{2015}]{ishimaru2015} Ishimaru, Y., Wanajo, S., \& Prantzos, N. 2015, ApJL, 804, L35
\bibitem[\protect\citeauthoryear{Ivans et al.}{2006}]{ivans2006} Ivans, I., Simmerer, J., Sneden, C., et al. 2006, ApJ, 645, 613
\bibitem[\protect\citeauthoryear{Johnson}{2002}]{johnson2002} Johnson, J. 2002, ApJS, 139, 219
\bibitem[\protect\citeauthoryear{Jacobson et al.}{2015}]{jacobson2015} Jacobson, H., Keller, S., Frebel, A., et al. 2015, ApJ, 807, 171
\bibitem[\protect\citeauthoryear{Ji et al.}{2016}]{ji2016} Ji, A. P., Frebel, A., Chiti, A., \& Simon, J. D. 2016, Natur, 531, 610
\bibitem[\protect\citeauthoryear{K\"{a}ppeler et al.}{1989}]{kappeler1989} K\"{a}ppeler, F., Beer, H., \& Wisshak, K. 1989, RPPh, 52, 945
\bibitem[\protect\citeauthoryear{Kobayashi et al.}{2006}]{kobayashi2006} Kobayashi, C., Umeda, H., Nomoto, K., et al. 2006, ApJ, 653, 1145
\bibitem[\protect\citeauthoryear{Komiya et al.}{2014}]{komiya2014} Komiya, Y., Yamada, S., Suda, T., \& Fujimoto, M. Y. 2014, ApJ, 783, 132
\bibitem[\protect\citeauthoryear{Komiya \& Shigeyama}{2016}]{komiya2016} Komiya, Y., \& Shigeyama, T. 2016, ApJ, 830, 76
\bibitem[\protect\citeauthoryear{Lamb et al.}{1977}]{lamb1977} Lamb, S., Howard, W. M., Truran, J. W., et al. 1977, ApJ, 217, 213
\bibitem[\protect\citeauthoryear{Lai et al.}{2008}]{lai2008} Lai, D. K., Bolte, M., Johnson, J. A., et al. 2008, ApJ, 681, 1524
\bibitem[\protect\citeauthoryear{Lattimer \& Schramm}{1974}]{lattimer1974} Lattimer, J. M., \& Schramm, D. N. 1974, ApJ, 192, L145
\bibitem[\protect\citeauthoryear{Li et al.}{2015a}]{li2015a} Li, H. N., Zhao, G., Christlieb, N., et al. 2015a, ApJ, 798, 110
\bibitem[\protect\citeauthoryear{Li et al.}{2015b}]{li2015b} Li, H. N., Aoki, W., Honda, S., et al. 2015b, RAA, 15, 1264L
\bibitem[\protect\citeauthoryear{Li et al.}{2013a}]{li2013a} Li, H. J., Cui, W. Y., \& Zhang, B. 2013a, ApJ, 775, 12
\bibitem[\protect\citeauthoryear{Li et al.}{2013b}]{li2013b} Li, H. J., Shen, X. J., Liang, S., et al. 2013b, PASP, 125, 143
\bibitem[\protect\citeauthoryear{Li et al.}{2014}]{li2014} Li, H. J., Ma, W. J., Cui, W. Y., et al. 2014, PASP, 126, 544
\bibitem[\protect\citeauthoryear{Machida et al.}{2005}]{machida2005} Machida, M. N., Tomisaka, K., Nakamura, F., \& Fujimoto, M. Y. 2005, ApJ, 622, 39
\bibitem[\protect\citeauthoryear{Mashonkina et al.}{2010}]{mashonkina2010} Mashonkina, L., Christlieb, N., Barklem, P. S., et al. 2010, A\&A, 516, 46
\bibitem[\protect\citeauthoryear{Mashonkina et al.}{2014}]{mashonkina2014} Mashonkina, L., Christlieb, N. \& Eriksson, K. 2014, A\&A, 569, A43
\bibitem[\protect\citeauthoryear{Mathews et al.}{1992}]{mathews1992} Mathews, G. J., Bazan, G., \& Cowan, J. J. 1992, ApJ, 391, 719
\bibitem[\protect\citeauthoryear{McWilliam et al.}{1995}]{mcwilliam1995} McWilliam, A., Preston, G. W., Sneden, C., et al. 1995, AJ, 109, 2757
\bibitem[\protect\citeauthoryear{McWilliam}{1998}]{mcwilliam1998} McWilliam, A. 1998, AJ, 115, 1640
\bibitem[\protect\citeauthoryear{Mishenina et al.}{2013}]{mishenina2013} Mishenina, T. V., Pignatari, M., Korotin, S. A., et al. 2013, A\&A, 552, A128
\bibitem[\protect\citeauthoryear{Montes et al.}{2007}]{montes2007} Montes, F., Beers, T. C., Cowan, J., et al. 2007, ApJ, 671, 1685
\bibitem[\protect\citeauthoryear{Nomoto et al.}{2013}]{nomoto2013} Nomoto, K., Kobayashi C. \& Tominaga, N. 2013, ARA\&A, 51, 457
\bibitem[\protect\citeauthoryear{Qian et al.}{1998}]{qian1998} Qian, Y. Z., Vogel, P., \& Wasserburg, G. J. 1998, ApJ, 494, 285
\bibitem[\protect\citeauthoryear{Qian \& Wasserburg}{2001}]{qian2001} Qian, Y. Z., \& Wasserburg, G. J. 2001, ApJ, 559, 925
\bibitem[\protect\citeauthoryear{Qian \& Wasserburg}{2007}]{qian2007} Qian, Y. Z., \& Wasserburg, G. J. 2007, PhR, 442, 237
\bibitem[\protect\citeauthoryear{Raiteri et al.}{1991}]{raiteri1991} Raiteri, C. M., Busso, M., Gallino, R., et al. 1991, ApJ, 367, 228
\bibitem[\protect\citeauthoryear{Raiteri et al.}{1993}]{raiteri1993} Raiteri, C. M., Gallino, R., Busso, M., et al. 1993, ApJ, 419, 207
\bibitem[\protect\citeauthoryear{Ramirez-Ruiz et al.}{2015}]{ramirez2015} Ramirez-Ruiz, E., Trenti, M., MacLeod, M., et al. 2015, ApJL, 802, L22
\bibitem[\protect\citeauthoryear{Roederer et al.}{2010}]{roederer2010} Roederer, I. U., Sneden, C., Thompson, I. B., et al. 2010, ApJ, 711, 573
\bibitem[\protect\citeauthoryear{Roederer et al.}{2014}]{roederer2014} Roederer, I. U., Preston, G. W., Thompson, I. B. 2014, AJ, 147, 136
\bibitem[\protect\citeauthoryear{Rosswog et al.}{2014}]{rosswog2014} Rosswog, S., Korobkin, O., Arcones, A., et al. 2014, MNRAS, 439, 744
\bibitem[\protect\citeauthoryear{Ryan et al.}{1996}]{ryan1996} Ryan, S. G., Norris, J. E., \& Beers, T. C. 1996, ApJ, 471, 254
\bibitem[\protect\citeauthoryear{Siqueira-Mello et al.}{2012}]{siqueira2012} Siqueira-Mello, C., Barbuy, B., Spite, M., \& Spite, F. 2012, A\&A, 548, A42
\bibitem[\protect\citeauthoryear{Siqueira-Mello et al.}{2016}]{siqueira2016} Siqueira-Mello, C., Chiappini, C., Barbuy, B., et al. 2016, A\&A, 593, A79
\bibitem[\protect\citeauthoryear{Sneden et al.}{1998}]{sneden1998} Sneden, C., Cowan, J. J., Burris, D. L., \& Truran, J. W. 1998, ApJ, 496, 235
\bibitem[\protect\citeauthoryear{Sneden et al.}{2003}]{sneden2003} Sneden, C., Cowan, J. J., Lawler, J. E., et al. 2003, ApJ, 591, 936
\bibitem[\protect\citeauthoryear{Sneden et al.}{2008}]{sneden2008} Sneden, C., Cowan, J. J., \& Gallino, R. 2008, ARA\&A, 46, 241
\bibitem[\protect\citeauthoryear{Sneden et al.}{1994}]{sneden1994} Sneden, C., Preston, G. W., McWilliam, A., \& Searle, L. 1994, ApJL, 431, L27
\bibitem[\protect\citeauthoryear{Travaglio et al.}{1999}]{travaglio1999} Travaglio, C., Galli, D., Gallino, R., et al. 1999, ApJ, 521, 691
\bibitem[\protect\citeauthoryear{Travaglio et al.}{2001}]{travaglio2001} Travaglio, C., Galli, D., \& Burkert, A. 2001, ApJ, 547, 217
\bibitem[\protect\citeauthoryear{Travaglio et al.}{2004}]{travaglio2004} Travaglio, C., Gallino, R., Arnone, E., et al. 2004, ApJ, 601, 864
\bibitem[\protect\citeauthoryear{Truran}{1981}]{truran1981} Truran, J. W. 1981, A\&A, 97, 391
\bibitem[\protect\citeauthoryear{Tsujimoto et al.}{1999}]{tsujimoto1999} Tsujimoto, T., Shigeyama, T., \& Yoshii, Y. 1999, ApJ, 519, L63
\bibitem[\protect\citeauthoryear{Tsujimoto \& Shigeyama}{2014a}]{tsujimoto2014a} Tsujimoto, T., \& Shigeyama, T. 2014a, ApJL, 795, L18
\bibitem[\protect\citeauthoryear{Tsujimoto \& Shigeyama}{2014b}]{tsujimoto2014b} Tsujimoto, T., \& Shigeyama, T. 2014b, A\&A, 565, L5
\bibitem[\protect\citeauthoryear{Wanajo et al.}{2003}]{wanajo2003} Wanajo, S., Tamamura, M., Itoh, N., et al. 2003, ApJ, 593, 968
\bibitem[\protect\citeauthoryear{Wasserburg et al.}{1996}]{wasserburg1996} Wasserburg, G. J., Busso, M., \& Gallino, R. 1996, ApJL, 466, L109
\bibitem[\protect\citeauthoryear{Westin et al.}{2000}]{westin2000} Westin, J., Sneden, C., Gustafsson, B., et al. 2000, ApJ, 530, 783
\bibitem[\protect\citeauthoryear{Woosley \& Weaver}{1995}]{woosley1995} Woosley, S. E., \& Weaver, T. A. 1995, ApJS, 101, 181
\bibitem[\protect\citeauthoryear{Yong et al.}{2013}]{yong2013} Yong, D., Norris, J. E., Bessell, M. S., et al. 2013, ApJ, 762, 26
\end{thebibliography}
\end{document}